\documentclass{llncs}
\usepackage{cite}
\usepackage{appendix}
\usepackage{graphicx}
\usepackage{amsmath}
\usepackage[byname]{smartref}
\usepackage{makeidx}
\usepackage{imakeidx}
\makeindex[columns=1]
\usepackage[bookmarks,pagebackref, pdfpagelabels=true, colorlinks=true, urlcolor=blue, citecolor = blue, anchorcolor = blue, linkcolor=blue, pdfborder={0 0 0}]{hyperref}
\usepackage{txfonts}
\usepackage{tocloft}
\usepackage{calrsfs}

\usepackage{algorithm}
\usepackage{algpseudocode}
\usepackage{adjustbox}
\usepackage{xspace}
\usepackage{mathrsfs}
\usepackage{breqn}
\usepackage{amsfonts, amssymb}
\usepackage{tikz}
\usepackage{tikzscale}
\usepackage{textcomp}
\usepackage{xcolor}
\usepackage{color}
\usepackage{sidecap}
\usepackage{graphicx}
\usepackage{pdfpages}
\usepackage{lipsum}
\usepackage{subcaption}

{\clearemptydoublepage 
 \begin{center}
  \section*{Dedication}
 \end{center}
 \begingroup
}{\newpage\endgroup}

{\pagestyle{plain}\pagenumbering{roman}}%
{\pagenumbering{arabic}}

\usepackage{color}
\newboolean{showComments}
\newcommand{\todocmd}[1]{\small{\textcolor{red}{#1}}}

\newcommand{\todo}[1]{\ifthenelse {\boolean{showComments}} {\todocmd{#1}} {}}

\numberwithin{equation}{section}

\newcommand{\maple}[1]{{\tt #1}\xspace}

\makeatother

\begin{document}

\title{Algorithmic reduction of polynomially nonlinear PDE systems to parametric ODE systems
}

\author{Siyuan Deng\inst{1,2} \and Michelle Hatzel\inst{1,3} \and
Gregory Reid\inst{1,4} \and Wenqiang Yang\inst{5,6} \and  Wenyuan Wu\inst{5,7}}

\institute{Department of Mathematics, Western University, London, Canada 
\and \email{sdeng53@uwo.ca}
\and \email{mhatzel@uwo.ca} 
\and \email{reid@uwo.ca} 
\and Chongqing Institute of Green and Intelligent Technology, Chinese Academy of Sciences, Chongqing, China \and \email{yangwenqiang@cigit.ac.cn} 
\and \email{wuwenyuan@cigit.ac.cn}}

\maketitle 

\begin{abstract}
Differential-elimination algorithms apply a finite number of differentiations and eliminations to systems 
of partial differential equations.  For systems that are polynomially nonlinear with rational number coefficients, they guarantee the inclusion of missing integrability conditions and the statement of 
of existence and uniqueness theorems for local analytic solutions of such systems.
Further, they are useful in obtaining systems in a form more amenable to exact and approximate solution methods. 

Maple's \maple{dsolve} and \maple{pdsolve} algorithms for solving 
PDE and ODE often automatically call such routines during applications.
Indeed even casual users of \maple{dsolve} and \maple{pdsolve} have probably unknowingly 
used Maple's differential-\\elimination algorithms.

Suppose that a system of PDE has been reduced by differential-\\elimination method
to a system whose automatic existence and uniqueness algorithm has been determined to 
be finite-dimensional.  We present an algorithm for rewriting the output as a system of 
parameterized ODE.
Exact methods and numerical methods for solving ODE and DAE can be applied to this form.

\keywords
    numerical analysis, partial differential equations, algebraic geometry, computer algebra, ordinary differential equations, DAE

\end{abstract}

\section{Introduction}

Maple has three powerful differential-elimination packages, including the RIF package \cite{RifsimpPackage},
the DifferentialAlgebra \cite{blop95:jet} package and the Differential Thomas package \cite{Robertz14:ThomasDecom}.

As an illustrative example, used throughout this article, consider the system of PDE $R$ given by:
\begin{equation}
\label{eq:simpEq}
\frac{\partial^{2}}{\partial x^{2}}u \! \left(x , y\right)-\frac{\partial^{2}}{\partial x \partial y}u \! \left(x , y\right)
 = 0, 
\left(\frac{\partial u(x,y) }{\partial y}\right  )^{2}  +\frac{\partial}{\partial y}u \! \left(x , y\right)-u \! \left(x , y\right)
 = 0
 \end{equation}
 We note that this is polynomially nonlinear with rational coefficients, so the above algorithms can be
 applied.  We also note that the system has two equations for one unknown function $u(x,y)$ and so is 
 over-determined. In particular, differentiating the first equation with respect to $y$ and the second equation with respect to $x$ will yield an integrability condition for the derivative $\frac{\partial^{3}}{\partial x^{2} \partial y}u $. This results in a non-trivial integrability 
 condition, and this system is one for which the above differential-elimination packages are useful.

 In the abstract, we claimed that most 
 casual users of Maple's \maple{dsolve} and \maple{pdsolve} had probably unknowingly also 
 used such differential-elimination packages.
 Indeed such users usually apply \maple{dsolve} and \maple{pdsolve} to a single differential equation,
 which are not over-determined and have no nontrivial integrability conditions seemingly
 contradicting our claim.

 However, consider the following ODE:
\begin{equation}
\label{eq:simpODE}
6 \left(\frac{d}{d x}y \! \left(x \right)\right) \left(\frac{d^{2}}{d x^{2}}y \! \left(x \right)\right)+2 y \! \left(x \right) \left(\frac{d^{3}}{d x^{3}}y \! \left(x \right)\right)+y \! \left(x \right)^{2}
 = 0
 \end{equation}
 
 When we use the Maple commands \maple{dsolve(DE)} and \maple{infolevel[rifsimp] := 4}, we see that multiple calls are 
 made the differential elimination routine \maple{rifsimp}.
 At first sight, this is puzzling since the DE is not over-determined and has no non-trivial 
 integrability conditions.

 However, an important class of integration methods for differential equations is based on finding infinitesimal Lie symmetry vector fields $\xi(x,y) \frac{\partial}{\partial x} + \eta(x,y)\frac{\partial}{\partial y} $ leaving the DE invariant.
 For background on symmetry methods see \cite{BlumanChevAnco2010} and \cite{Olv93:App}.
 Applying the Maple command \maple{[PDEtools]DetermingPDE} with the option \maple{integrabilityconditions := false} to the DE yields an over-determined system of 9 PDE 
 for $\xi, \eta$, the coefficients of the symmetry vector fields leaving the DE invariant.
 The first 4 (shortest) equations of that system are:
 \begin{equation}
 \label{eq:simpODEdeteqs}
\left[ \frac{\partial}{\partial y} \xi  = 0, \; \; 
\frac{\partial^{2}}{\partial y^{2}}\xi  = 0, \; \;
2 y \frac{\partial^{3}}{\partial y^{3}}\xi  
+6 \frac{\partial^{2}}{\partial y^{2}}\xi 
 = 0,   
6 y \frac{\partial^{2}}{\partial x^{2}}\xi
-6 y \frac{\partial^{2}}{\partial x \partial y}\eta
-6 \frac{\partial}{\partial x}\eta 
 = 0,  \cdots \right]
 \end{equation}

 Such symmetry determining systems are linear in their coefficients, and usually over-determined.
 Differential-elimination methods are natural for such problems 
 and have proved to be standard tools for simplifying such 
 systems. Applying \maple{rifsimp} to the above system yields RIF form given by:
 \begin{equation}
 \label{eq:RIF-ODEdetsys}
 \left[
 \frac{\partial^{3}}{\partial x^{3}}\eta =\frac{y\left(\frac{\partial}{\partial y}\eta \right)}{2}-\frac{\eta }{2}, 
 \frac{\partial^{2}}{\partial y \partial x}\eta =-\frac{\frac{\partial}{\partial x}\eta }{y}, \frac{\partial^{2}}{\partial y^{2}}\eta =\frac{- y \left(\frac{\partial}{\partial y}\eta \! \right)  +\eta }{y^{2}}, \frac{\partial}{\partial x}\xi =0, \frac{\partial}{\partial y}\xi =0
\right]
 \end{equation}
The \maple{initialdata} algorithm yields that the solution space is finite-dimensional with 5 dimensional
initial data given by
\begin{equation}
\label{eq:ID-ODEdetsys}
    \left[\eta \! \left(x_{0}, y_{0}\right)=C_1 , \eta_x \left(x_{0}, y_{0}\right)=C_2 , \eta_y \left(x_{0}, y_{0}\right)=C_3 , \eta_{xx} \left(x_{0}, y_{0}\right)=C_4 , \xi \! \left(x_{0}, y_{0}\right)=C_5 \right]
\end{equation}
Additionally, the Lie Algebra of vector fields package (LAVF) enables the structure of the 5-dimensional Lie symmetry algebra to be computed directly from the RIF-form and the initial data. The Lie algebra structure is:
\begin{eqnarray*}
[X_{1}, X_{2}] &=& \frac{X_{5}}{2}, [X_{1}, X_{3}] = 
X_{1}-\frac{X_{3}}{y}, [X_{1}, X_{4}] = -\frac{X_{4}}{y}, [X_{1}, X_{5}] = -\frac{X_{5}}{y}, [X_{2}, X_{3}] = \frac{y X_{5}}{2},    \\    
\phantom{x}  [X_{2}, X_{4}] &=& X_{1}-\frac{X_{3}}{y}, [X_{2}, X_{5}] = X_{4}, [X_{3}, X_{4}] = -X_{4}, [X_{3}, X_{5}] = -X_{5} 
\end{eqnarray*}
where $y = y_0 \not = 0$ is a constant. Here we can use the commands to determine the dimension of the 
the derived algebra $\dim DerivedAlgebra (L) = 3$.
In particular an $r$-th order ODE is linearizable iff $\dim DerivedAlgebra(L) = r$
and $DerivedAlgebra(L)$ is abelian.
Applying these results and the algorithms given in Mohammadi, Reid and Huang\cite{MRH19} and 
Lyakhov, Gerdt and Michels\cite{LyakhovGerdtMichels17}
shows that the ODE is linearizable and with linearizing transformation 
$\hat{u} = u^2$, $\hat{x} = x$ and target linear ODE 
$\left( \frac{d}{d \hat{x}} \right)^3 \hat{u} + \hat{u} = 0$.

\section{Reduction of systems of PDE with finite-dimensional solution spaces to parameterized ODE}

The defining systems for symmetries of ODE and PDE are often over-determined as discussed in the introduction.  Consequently as discussed in the introduction, differential-elimination algorithms
have become essential tools for the determination of symmetries and mappings of differential equations.
Such algorithms are also important in the analysis of differential equations with constraints (so-called
DAE) which arise naturally from modeling environments such as \maple{MapleSim} and \maple{SystemModeler}.


For an algorithmic approach
we need to exploit the algorithmic existence and uniqueness results from differential-elimination 
algorithms for polynomially nonlinear differential systems with rational coefficients.
Here we exploit the results for the rifsimp algorithm. See Rust, Reid and Wittkopf\cite{RGW99}
for details of the existence and uniqueness results.

We now give a brief outline of our algorithm and its justification for reducing systems of differential
equations with finite-dimensional solution spaces. A more detailed exposition will be given elsewher.
Let $R$ denote an exact system of polynomially nonlinear PDE with independent variables $x$ and dependent variables $u$ and rational coefficients and let $\prec$ be a ranking of derivatives \cite{Rust:Thesis}.

Given $\prec$ the RIF algorithm applies a finite number of differentiations and eliminations to $R$ outputting a finite number of cases labeled by $j$
with an associated local existence and uniqueness theorem.
Each case consists of a system of equations $E_j$, and a system of inequations $I_j$.
Let $Prin (E_j)$ be the leading derivatives of $E_j$ with respect to $\prec$.
Then the number of free parameters in solutions of $E_j$ is finite, it is determined locally near $x^0$ by a finite list of parametric derivatives of $u$. 
Considering this list as new dependent variables $v$ for $R_j$ near $x^0$, $R_j$ can be rewritten in first-order form 
$\frac{\partial}{\partial x_i} v = f_i (x, v)$, $h(x, v) = 0 $.  Thus, $R$ is rewritten as a system of 
parametrized ODE on a constraint.

\vspace{2cm} 

\noindent
\textbf{Example 1} 

If we apply the RIF algorithm to $R$, then provided
$2 \frac{\partial}{\partial y} u(x , y)+1 \not = 0$,
the leading linear system of RIF($R$) is
\begin{align*}
     \mathit{LeadingLinear} =\left[\frac{\partial^{2}}{\partial x^{2}}u \! \left(x , y\right)=\right. &\left.\frac{\frac{\partial}{\partial x}u \! \left(x , y\right)}{2 \frac{\partial}{\partial y}u \! \left(x , y\right)+1}, \frac{\partial^{2}}{\partial y \partial x}u \! \left(x , y\right)=\frac{\frac{\partial}{\partial x}u \! \left(x , y\right)}{2 \frac{\partial}{\partial y}u \! \left(x , y\right)+1}, \right. \\ &\left. \frac{\partial^{2}}{\partial y^{2}}u \! \left(x , y\right)=\frac{\frac{\partial}{\partial y}u \! \left(x , y\right)}{2 \frac{\partial}{\partial y}u \! \left(x , y\right)+1}\right],
\end{align*}
and the leading nonlinear system of RIF($R$) is
\begin{align*}
    \mathit{LeadingNonlinear} =&\left[\left(\frac{\partial}{\partial x}u \! \left(x , y\right)\right) \left(\frac{\partial}{\partial y}u \! \left(x , y\right)\right)-\left(\frac{\partial}{\partial x}u \! \left(x , y\right)\right)^{2}=0,\right. \\ &\left. \left(\frac{\partial}{\partial y}u \! \left(x , y\right)\right)^{2}+\frac{\partial}{\partial y}u \! \left(x , y\right)-u \! \left(x , y\right)=0\right].
\end{align*}

Note that RIF also computes all possible splitting on coefficients of the leading linear system. Here, that yielded two cases 
$2 \frac{\partial}{\partial y} u(x , y)+1 \not = 0$
and $2 \frac{\partial}{\partial y} u(x , y)+1  = 0$.
The second leads to a branch with no solutions, so that case
is discarded.

The key to decoupling the RIF form above into $x$ derivatives and $y$ derivatives is to compute the 
parametric derivatives of the leading linear PDE.
First, the leading derivatives of the leading linear PDE are
computed, yielding 
$$
 \mathit{Leading Derivatives} =\left[\frac{\partial^{2}}{\partial x^{2}}u \! \left(x , y\right), \frac{\partial^{2}}{\partial y \partial x}u \! \left(x , y\right), \frac{\partial^{2}}{\partial y^{2}}u \! \left(x , y\right)\right].
$$

Next, the complementary set of parametric derivatives is computed.  These 
are all the derivatives that are not derivatives of the 
leading derivatives.  They are:
$$  {\cal P} = \left[ u \! \left(x , y\right),   \frac{\partial}{\partial x} u \! \left(x , y\right), 
 \frac{\partial}{\partial y} u \! \left(x , y\right)
 \right].
 $$ 
 
 Relabelling these parametric derivatives as new dependent 
 variables yields
 $$ \left[ u = u\left(x , y\right), u_x =   \frac{\partial}{\partial x} u \! \left(x , y\right), 
 u_y = \frac{\partial}{\partial y} u \! \left(x , y\right)
 \right]. $$
 
 Notice for simplicity of notation; we have used $u_x$ and $u_y$ to denote new dependent variables.
 Computing RIF form with respect to the original system
 yields the ODE system with respect to $x$:
 $$\left[\frac{\partial}{\partial x} u = u_{x}, \frac{\partial}{\partial x} u_x = \frac{u_{x}}{2 u_{y}+1}
, 
\frac{\partial}{\partial x} u_{y} = \frac{u_{x}}{2 u_{y}+1} \right], $$

and the ODE system with respect to $y$:
$$
\left[\frac{\partial}{\partial y} u = u_{y}, 
\frac{\partial}{\partial y} u_{x}   = \frac{u_{x}}{2 u_{y}+1}, 
\frac{\partial}{\partial y} u_{y}   = \frac{u_{y}}{2 u_{y}+1}\right],
$$

where these ODE with respect to $x$ and $y$ share the constraint:
$$ u_x u_y - u_x^2 = 0 ,$$

and the inequation $2 u_y + 1 \not = 0$.


\noindent
\textbf{Example 2}

We apply the algorithm to the RIF-form determining system given in 
(\ref{eq:RIF-ODEdetsys}) and corresponding finite-dimensional initial data given
in (\ref{eq:ID-ODEdetsys}).
The list of parametric derivatives is 
$$ \xi, \eta , \eta_x , \eta_y , \eta_{xx} $$
Then taking $x$ derivatives of this list and simplifying with respect to the RIF form
(\ref{eq:RIF-ODEdetsys}) yields the system of ODE with respect to $x$ with $y$ regarded as
a parameter:
\begin{equation}
   \left[ \frac{\partial}{\partial x} \xi = 0 , \frac{\partial}{\partial x} \eta = \eta_x, 
    \frac{\partial}{\partial x} \eta_x = \eta_{xx}, 
    \frac{\partial}{\partial x} \eta_y = -\frac{\eta_x }{y},
    \frac{\partial}{\partial x} \eta_{xx} = \frac{y \eta_y }{2}-\frac{\eta }{2} \right]
\end{equation}
Similarly taking $y$ derivatives of this list and simplifying with respect to the RIF form
(\ref{eq:RIF-ODEdetsys}) yields the system of ODE with respect to $y$ with $x$ regarded as a parameter:
\begin{equation}
 \left[ \frac{\partial}{\partial y} \xi = 0 , \frac{\partial}{\partial y} \eta = \eta_y, 
    \frac{\partial}{\partial y} \eta_x = -\frac{\eta_x }{y}, 
    \frac{\partial}{\partial y} \eta_y = \frac{- y \eta_y  +\eta }{y^{2}},
    \frac{\partial}{\partial y} \eta_{xx} =  -\frac{\eta_{xx} }{y}  \right]
\end{equation}


\section{Discussion and Applications}

The above computations have been automated in the RIF package available in distributed Maple, and could also be 
implemented in other symbolic differential elimination packages.

\subsection{Analytical solutions}  
Reduction of PDE to ODE has obvious advantages because of the breadth of available ODE methods.  Indeed, early examples of such computations date back to classical differential geometers and, in particular, Cartan \cite{Cartan:sym}, whose exterior differential systems naturally express the ODE-like character of such overdetermined PDE systems.

\subsection{Numerical solutions for exact polynomially nonlinear PDE}

The numerical solution of such PDE systems for exact polynomially nonlinear PDE, via the reduction
method we outlined above, involves the solution of ordinary differential equations on manifolds (or so-called Differential Algebraic Equation - DAE). The RIF algorithm reduces the involved DAE to the index one case. A point with the initial data $v(x^0) = v^0$ determines a unique local solution.
Then a DAE solver can approximate this solution along a curve through this point.
Points on this solution curve can then be used as initial values for the DAE to approximate solutions
and build, by iteration, an approximation of the local solution to $v(x^0) = v^0$.
Details will be given elsewhere.

\subsection{Numerical Polynomial Algebra}

A key area of the conference is polynomial and 
matrix algebra.  Indeed suppose that we are considering 
polynomial systems in a ring $\mathbb{Q}$ in the indeterminates $x_1, x_2, \cdots, x_n$.
Then ideal computations in the polynomial ring 
$\mathbb{Q} [x_1 , \cdots , x_n ]$ could equivalently be done
in the differential ring $\mathbb{Q} [\frac{\partial}{\partial x_1} , \cdots , \frac{\partial}{\partial x_n} ]$ 
via the map $x_j \leftrightarrow \frac{\partial}{\partial x_j}$.  Indeed, using this mapping, the methods described in our article correspond to eigenvalue-matrix methods for solving $0$-dimensional polynomial systems.

For example, Michałek and Sturmfels
\cite{MichałekSturmfels21} consider the polynomial ideal
$\langle x^3 - y z, y^3 - x z, z^3 - x y \rangle$
and associate this with the differential ideal
$\langle \left( {\frac{\partial}{\partial x}}\right)^3 -  {\frac{\partial}{\partial y}} {\frac{\partial}{\partial z}}, \left( {\frac{\partial}{\partial y}}\right)^3 - {\frac{\partial}{\partial x}} {\frac{\partial}{\partial z}}, \left( {\frac{\partial}{\partial z}}\right)^3 - {\frac{\partial}{\partial x}} {\frac{\partial}{\partial y}} \rangle$. Following the algorithm of our article, yields a 27-dimensional ideal, and yields 3 ODE systems of the form:
$$ \frac{\partial}{\partial x} \textbf{v} = X \textbf{v}, 
   \frac{\partial}{\partial y} \textbf{v} = Y \textbf{v},
   \frac{\partial}{\partial z} \textbf{v} = Z \textbf{v} $$
   where $X, Y, Z$ are sparse $ 27 \times 27$ matrices that mutually commute, and $\textbf{v}$ is a \\$27$-dimensional vector. Solving these elementary ODE systems yields the same result as in the text \cite{MichałekSturmfels21}. As the authors describe the different representations (here differential versus polynomial) can yield valuable insights.

\subsection{Numerical solutions for approximate polynomially nonlinear PDE}

Many applications for PDE involve approximate parameters. 
Thus applying exact \\differential-elimination algorithms such as the RIF algorithm, can be subject to issues such as pivoting on small coefficients.  The strategy we outlined above enables the solution in terms of ODE prolongations.  It is natural to ask whether we can exploit such ODE prolongations without using the RIF algorithm.  Indeed we can take a system of PDE, and for example, substitute derivatives up to some given order as new dependent variables.  Then, potentially, the ODE prolongation with respect to each independent variable could be computed using methods such as those of Pantiledes\cite{P88}, Pryce\cite{P98,P01} and Yang, Wu and Reid\cite{WYG21}.  If all these prolongations are finite-dimensional, then numerically, a bound can be found for in which to search for additional integrability conditions, in an incremental way, without the strong ordered (and unstable) elimination of the exact methods.
Reduction to ODE on constraints also potentially enables more efficient prolongation methods
to be developed for approximate systems of PDE, based on ODE prolongation structures.

\section*{Acknowledgements}
We thank the referees for their helpful comments.

\end{document}